\documentclass{aa}

\usepackage{xcolor}
\usepackage{graphicx}
\usepackage{txfonts}
\usepackage[breaklinks,colorlinks,citecolor=blue]{hyperref}

\usepackage{natbib}
\bibpunct{(}{)}{;}{a}{}{,} % to follow the A&A style

\definecolor{bittersweet}{rgb}{1.0, 0.44, 0.37}

\def\be{\begin{equation}}
\def\ee{\end{equation}}

\definecolor{dgreen}{rgb}{0, 0.6, 0.0}

\begin{document}

   \title{Biased tracers as a probe of beyond-$\Lambda$CDM cosmologies}

%   \subtitle{I. Overviewing the $\kappa$-mechanism}

   \author{Farbod Hassani\inst{1,2} \fnmsep\thanks{\email{farbod.hassani@astro.uio.no}}\and
          Julian Adamek \inst{3}
           \and
          Ruth Durrer \inst{4}
            \and
          Martin Kunz \inst{4}
          }

   \institute{Institute of Theoretical Astrophysics, Universitetet i Oslo, P.O. Box 1029 Blindern, N-0315 Oslo, Norway
   \and 
   Department of Physics, McGill University, 3600 rue University, Montreal QC H3A 2T8, Canada
    \and
     Institute for Computational Science, Universit\"at Z\"urich, Winterthurerstrasse 190, 8057 Z\"urich, Switzerland
     \and
     D\'epartement de Physique Th\'eorique, Universit\'e de Gen\`eve,
24 quai Ernest-Ansermet, 1211 Gen\`eve 4, Switzerland}

%   \date{Received September 15, 1996; accepted March 16, 1997}

% \abstract{}{}{}{}{} 

\abstract
{
Cosmological models beyond $\Lambda$CDM, such as those featuring massive neutrinos or modifications of gravity, often display a characteristic change (scale-dependent suppression or enhancement) in the matter power spectrum when compared to a six-parameter $\Lambda$CDM baseline. It is therefore a widely held view that constraints on those models can be obtained by searching for such features in the clustering statistics of large-scale structure. However, when using biased tracers of matter in the analysis, the situation is complicated by the fact that the bias also depends on cosmology. Here we investigate how the selection of tracers affects the observed signatures for two examples of beyond-$\Lambda$CDM cosmologies: massive neutrinos and clustering dark energy ($k$-essence). We study the signatures in the monopole, quadrupole, and hexadecapole of the redshift-space power spectra for halo catalogues from large $N$-body simulations and argue that a fixed selection criterion based on local attributes, such as tracer mass, leads to a near loss of signal in most cases. Instead, the full signal is recovered only if the selection of tracers is done at fixed bias. This emphasises the need to model or measure the bias parameters accurately in order to get meaningful constraints on the cosmological model.
}


   \maketitle
%
%________________________________________________________________

\section{Introduction}
The six-parameter $\Lambda$-cold-dark-matter ($\Lambda$CDM) model of cosmology fits most current cosmological observations~\citep{SupernovaSearchTeam:1998fmf, SupernovaCosmologyProject:1998vns,BOSS:2016wmc,Planck:2018vyg} but it relies on two unexplained ingredients, 
 dark matter (DM) and dark energy ($\Lambda$), which make up about $95 \%$ of the energy budget of the  Universe. 
 For this reason, cosmologists are investigating alternative models of dark energy and DM and also theories of gravity which deviate from general relativity on large scales. In the coming years, the next generation of large-scale structure (LSS) surveys  \citep{Amendola:2016saw,EUCLID:2011zbd,LSSTScience:2009jmu} will reach an extraordinary statistical power that will allow us to discriminate between different theories and possibly reveal the nature of the dark components. These surveys will map the three-dimensional galaxy distribution up to redshift $z\simeq 2$. Under the assumption that galaxies and their host halos trace the underlying DM distribution, we can measure the three-dimensional mass distribution in the Universe.

% Importance of modeling.
To keep up with these future observations we need sufficiently accurate theoretical predictions for the models we wish to study. 
While perturbation theory is sufficient to model the anisotropies of the cosmic microwave background (CMB), the galaxy distribution is more complicated and must be modelled with 
cosmological $N$-body simulations that also provide
access to a large number of modes  in the non-linear regime \citep{Angulo:2021kes}. As a result, $N$-body simulations of so-called non-standard cosmologies have been developed over recent years \citep{Baldi:2008ay,Barreira:2013eea,Llinares:2013jza,Adamek:2017uiq,LiBook,Hassani:2019lmy,Hassani:2020rxd}. While far from exhausting all possibilities, these help us to place stringent constraints on modified theories of gravity or models of dark energy 
and DM by comparing the details of structure formation in the different scenarios.

% our paper
At the most basic level, such a comparison often starts with the matter power spectrum. The different scenarios typically lead to a scale-dependent and redshift-dependent modification of the power spectrum when compared to a $\Lambda$CDM baseline, and this signature can then be used to constrain the models. When applying this reasoning to observed clustering in LSS, which is always measured from biased tracers of matter such as galaxies, one needs to keep in mind that the bias of a fixed type of tracer is not independent of cosmology either. In the simplest halo bias model \citep{Kaiser:1984sw}, the linear bias parameter for a fixed halo mass is approximately inversely proportional to the amplitude of fluctuations. This is because it becomes proportionally more difficult to form objects of a given mass when the initial fluctuations are smaller. Consequently, for a fixed type of halo defined through a mass threshold, a change in matter power is expected to be partially compensated by a change in bias, such that the clustering of the tracer is less affected by a change in cosmology.

Here, we present a study of this issue in the context of two beyond-$\Lambda$CDM scenarios: a cosmology with massive neutrinos and a cosmology with $k$-essence-type clustering dark energy. In the first case, the matter power gets suppressed on small scales due to the free streaming of neutrinos that make up a certain fraction of the matter as determined by their mass. In the second case, the growth rate gets modified at all scales due to a change in the expansion history of the late Universe. Additionally, the clustering of dark energy leaves a certain scale-dependent imprint. We use large $N$-body simulations to produce halo catalogues for the different scenarios and show that the model-specific signatures get almost completely wiped out in the halo power spectra if the comparison is carried out using halo populations with a fixed mass threshold. We further show, rather
trivially, that the expected signal is recovered if the comparison is instead carried out at fixed bias value. What is perhaps less trivial is that these statements also apply for the multipoles of the power spectra in redshift space, and we demonstrate this for the monopole, quadrupole, and hexadecapole.
   
%__________________________________________________________________
\section{Theory}
\label{sec:theory}

\begin{figure}[tb]%
    \centering
    \hspace*{-1cm}  
   {{\includegraphics[scale=0.33]{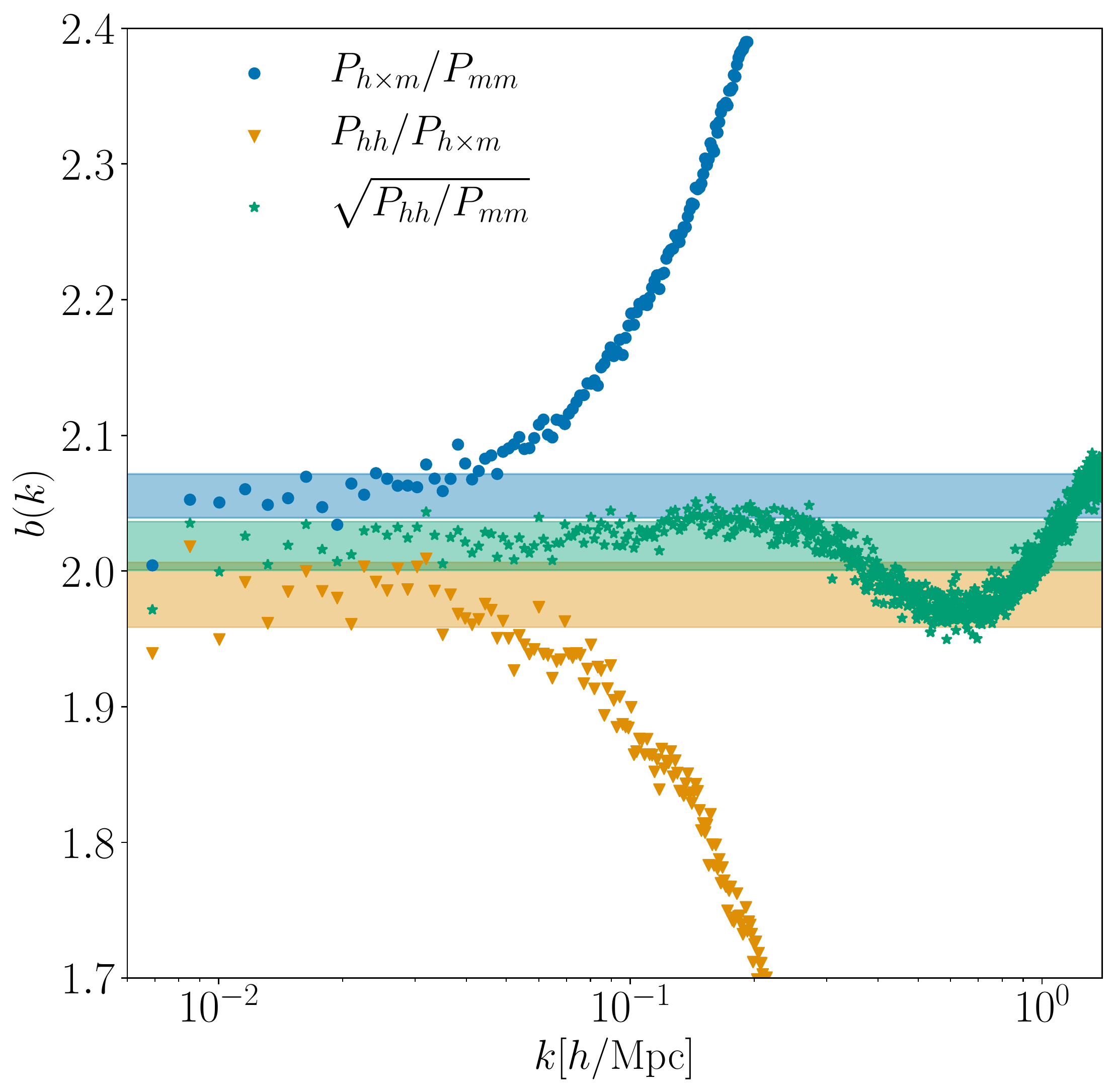} }}%
        \caption{ Halo bias obtained at redshift $z=0$ from three different definitions for a $\Lambda$CDM simulation with box size $4032$ Mpc/$h$ and $N_p= 4608^3$ particles. The bias is computed for halos with mass $M_h \ge 2 \times 10^{13}$\({M}_\odot\)/$h$. Bands show the confidence intervals for the linear bias $b_1$.}
    \label{bias}%
\end{figure}

We define the halo bias at a given wave number 
as
\be
b(k) = b_1 + b_\mathrm{NL}(k) = \sqrt{\frac{P_{hh}(k) }{P_{mm}(k)}}\,, \label{bias_def}
\ee
where $P_{hh}(k) $ and $P_{mm}(k)$ are, respectively, the 
halo and matter density power spectra without redshift-space distortions (RSDs), and $b_1$ is the linear bias parameter that captures the bias at large scales where matter and halo distributions are maximally correlated. At these scales, the non-linear bias $b_\mathrm{NL}(k) \to 0$. The bias $b(k)$ is a measure of how well the halo density field traces the underlying matter density  at a given scale; see \cite{Desjacques:2016bnm} for a review.  Alternative definitions would be $b(k) = P_{mh}(k) /P_{mm}(k)$ or $b(k) = P_{hh}(k) /P_{mh}(k)$, where $P_{mh}(k)$ is the matter-halo cross power spectrum. We note that in order to suppress the shot noise in our halo power spectra we employ a simple jack-knife method as described for example\ in \citet{Inman:2015pfa}. In Fig. \ref{bias} we show the bias as a function of wavenumber obtained from these different definitions for a $\Lambda$CDM simulation. While the non-linear bias depends on the definition because of the non-trivial correlation between matter and halos on small scales, the linear bias is almost the same within statistical error bars, for all definitions. In our simulations, we measure the linear bias 
by taking averages over $b(k)$ %(in Eq.~\ref{bias_def}) 
on modes in the linear and quasi-linear regime.

%%% bias

We consider two different selection procedures for the halo populations. In the first case, the population is selected according to a fixed mass threshold. We should note that our particle-mesh $N$-body simulations have a fixed effective force resolution (no adaptive mesh refinement) and the measured mass of the halos is not expected to be fully converged numerically. The quoted value of the mass threshold should therefore not be taken too literally, but rather considered a mass proxy. The important point is that the force resolution was not changed between runs of different cosmologies, and hence the mass proxies can be considered comparable. This selection method therefore selects, in some specific sense, the same types of objects across different models.

In the second case, we want to select the halo populations at a fixed value of the linear bias. This is achieved by measuring the bias for several choices of mass threshold, and then fitting for the threshold value that provides the desired bias. This selection procedure guarantees that any modifications of the matter power spectrum will appear in the corresponding halo power spectrum on all scales where the linear bias model is accurate. We note that carrying out a similar selection in a catalogue from an actual LSS survey would require some procedure for measuring the bias from observations.

\section{Non-standard cosmologies}
The two non-standard scenarios we consider in this work are 
massive neutrino cosmology and clustering dark energy ($k$-essence). 
In both cases, we use existing simulation suites that include suitable $\Lambda$CDM reference runs such that the impact of the non-standard scenarios on the clustering statistics can be quantified. To minimise the contamination from cosmic variance, the simulations are initialised on the same realisation of the Gaussian random field that sets the initial perturbations.

\paragraph{Massive neutrinos}
Laboratory measurements of neutrino flavour oscillations show that neutrinos have mass \citep{Esteban:2018azc}.
However, the absolute neutrino mass scale has not yet been determined, but only mass differences are known. It is expected that we will obtain the strongest constraints on the sum of neutrino masses  through cosmological probes \citep{Lesgourgues:2006nd}. This motivates cosmologists to study the impact of massive neutrinos on LSS through $N$-body simulations. In this work we use a suite of large $N$-body simulations from \citet{Adamek:2017uiq} that include neutrinos with the sum of masses given by $0$, $0.06$, $0.2,$ and $0.3$\,eV.
These simulations have a box size of $2048$ Mpc/$h$ with $4096^3$ particles, corresponding to a spatial resolution of $0.5$ Mpc/$h$ and mass resolution of about $ 10^{10} $\({M}_\odot\)/$h$. %\fhc{In the Rockstar output the particle mass is: 8.65801e+10 Msun/h. Should I be worried about it? When I compute the mass resolution myself I get $10^{10}$ which is the same as what is reported in the Neutrino paper.}
The $\Lambda$CDM baseline model in this case is the one with vanishing neutrino masses.

\paragraph{$k$-essence dark energy}
The $k$-essence model is a viable candidate for the late-time accelerated expansion of the Universe. This model was first introduced in \cite{ArmendarizPicon:2000ah} to avoid problems such as fine tuning, coincidence, or anthropic reasoning. 
The expansion history is slightly different from $\Lambda$CDM at low redshift due to the fact that $k$-essence has an effective equation of state of $w \neq -1$.
Moreover, the $k$-essence field clusters 
around matter over-densities depending on its speed of sound. This model was studied thoroughly by \citet{Hassani:2019lmy, Hassani:2019wed, Hassani:2020buk} using cosmological $N$-body simulations based on the code $k-$evolution. %~\citep{Hassani:2019lmy}.
Here we use two of these simulations with a fixed equation of state parameter $w=-0.9$ and two choices of the speed of sound, $c_s^2 = 1 $ and $c_s^2 = 10^{-4}$. The simulations have a box size of $4032$ Mpc/$h$ with $4608^3$ particles, corresponding to a spatial resolution of $0.875$ Mpc/$h$ and 
mass resolution of about $6 \times 10^{10} $\({M}_\odot\)/$h$. 

\section{Results}
In this section, we discuss the signatures of the $k$-essence and massive neutrino models in the
matter and halo power spectra. We study the even moments of the spectra in redshift space, that is,\ the monopole, quadrupole, and hexadecapole, which contain  the most information about clustering (including bias) and standard RSDs on linear and quasi-linear scales \citep{Kaiser:1987qv}.
The spectra are computed using the code \textit{Pylians3}\footnote{\url{https://pylians3.readthedocs.io/en/master/}} on the 
snapshots at $z=0$, and Doppler RSD are included using the distant-observer approximation where the velocity-induced shifts in redshift are parallel to one of the coordinate axes. Our halo catalogues are produced with the \textit{Rockstar} halo finder \citep{Behroozi:2011ju}. From each halo catalogue, we select two samples as explained in Section \ref{sec:theory}: one using a fixed mass threshold, and one using a fixed bias value by adjusting the mass threshold accordingly. For the matter power spectra, as well as for the halo power spectra from each selection method, we compute the ratios with respect to the one found in the corresponding $\Lambda$CDM simulation. As the initial conditions were chosen to be perfectly correlated across simulations, cosmic variance largely drops out in these ratios.
      
\begin{figure}[tb]%
    \centering
    \hspace*{-1cm}  
   {{\includegraphics[scale=0.33]{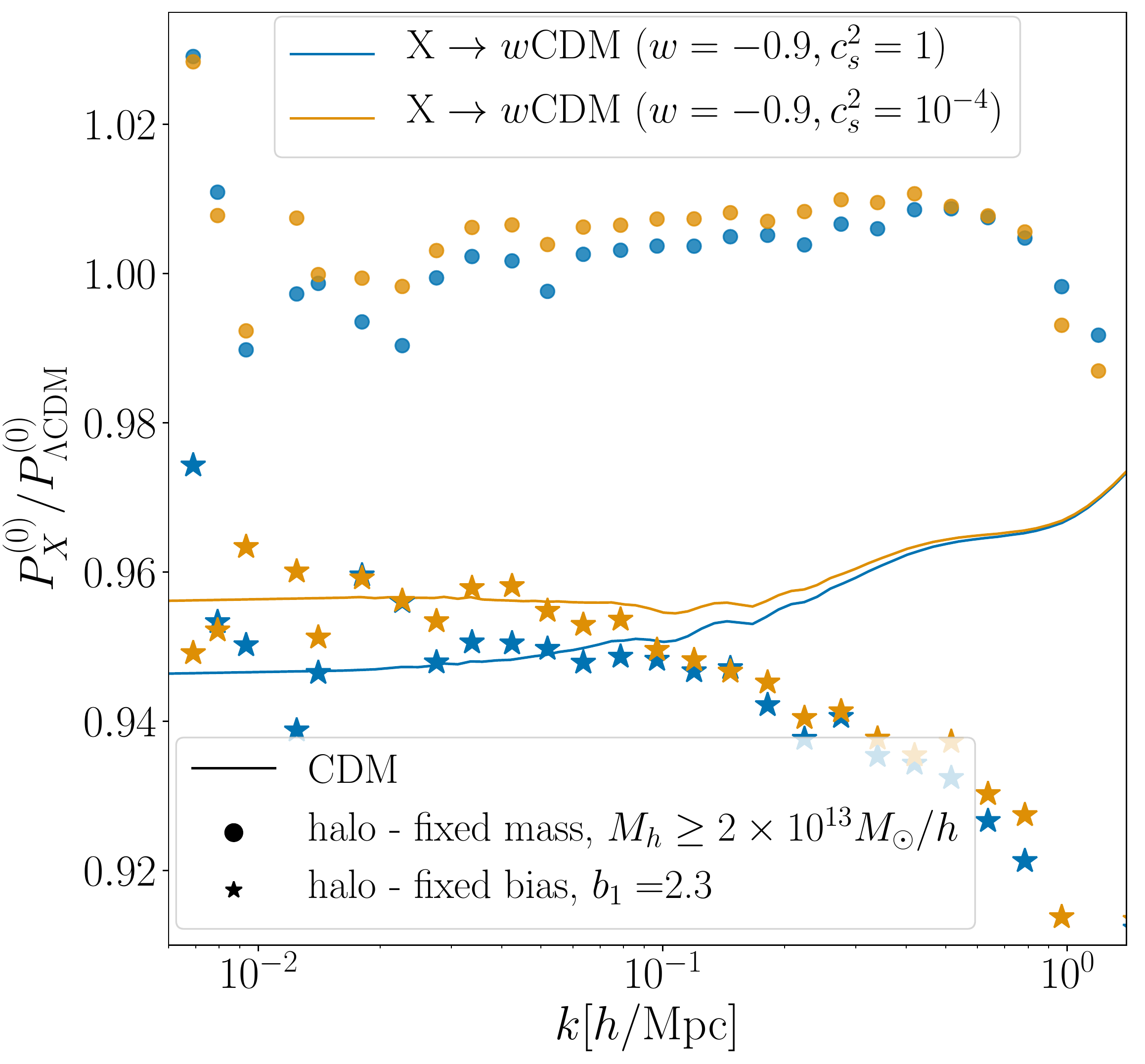} }}%
        \caption{ Ratio of $k$-essence to $\Lambda$CDM halo monopole power spectrum  (symbols) compared to the ratio of the corresponding matter power spectra (solid lines). The stars represent the result for the fixed bias selection, while the circles  correspond to the selection at fixed mass threshold.} 
    \label{kessence_pow}%
\end{figure}

In Fig.~\ref{kessence_pow} we show the results for the monopole power spectra from the $k$-essence simulations. Blue colour is used to show the case where $c_s^2 = 1$ and orange colour is used for the case where $c_s^2 = 10^{-4}$. 
%Solid lines %The lines
%represent the ratios for the matter power spectra, while the stars and circles show the ratios for the halo power spectra.
The stars correspond to the samples selected at fixed bias while the circles are for  halos with fixed mass threshold.
The figure shows clearly that the first selection can recover the signature in the monopole of the power spectrum very well in the linear and quasi-linear regime, $k\lesssim 0.1 h$/Mpc. On the other hand, in the power spectra of the halos selected according to fixed mass threshold, the signal is completely removed and the ratio of the spectra is close to unity. We observe a similar behaviour for the other moments of the power spectrum for the $k-$essence models, which, on the other hand, is less visible due to the weakness of the effect and to the large noise in the higher multipoles.

%%% monopole_kess

\begin{figure}[tb]%
    \centering
    \hspace*{-1cm}  
   {{\includegraphics[scale=0.33]{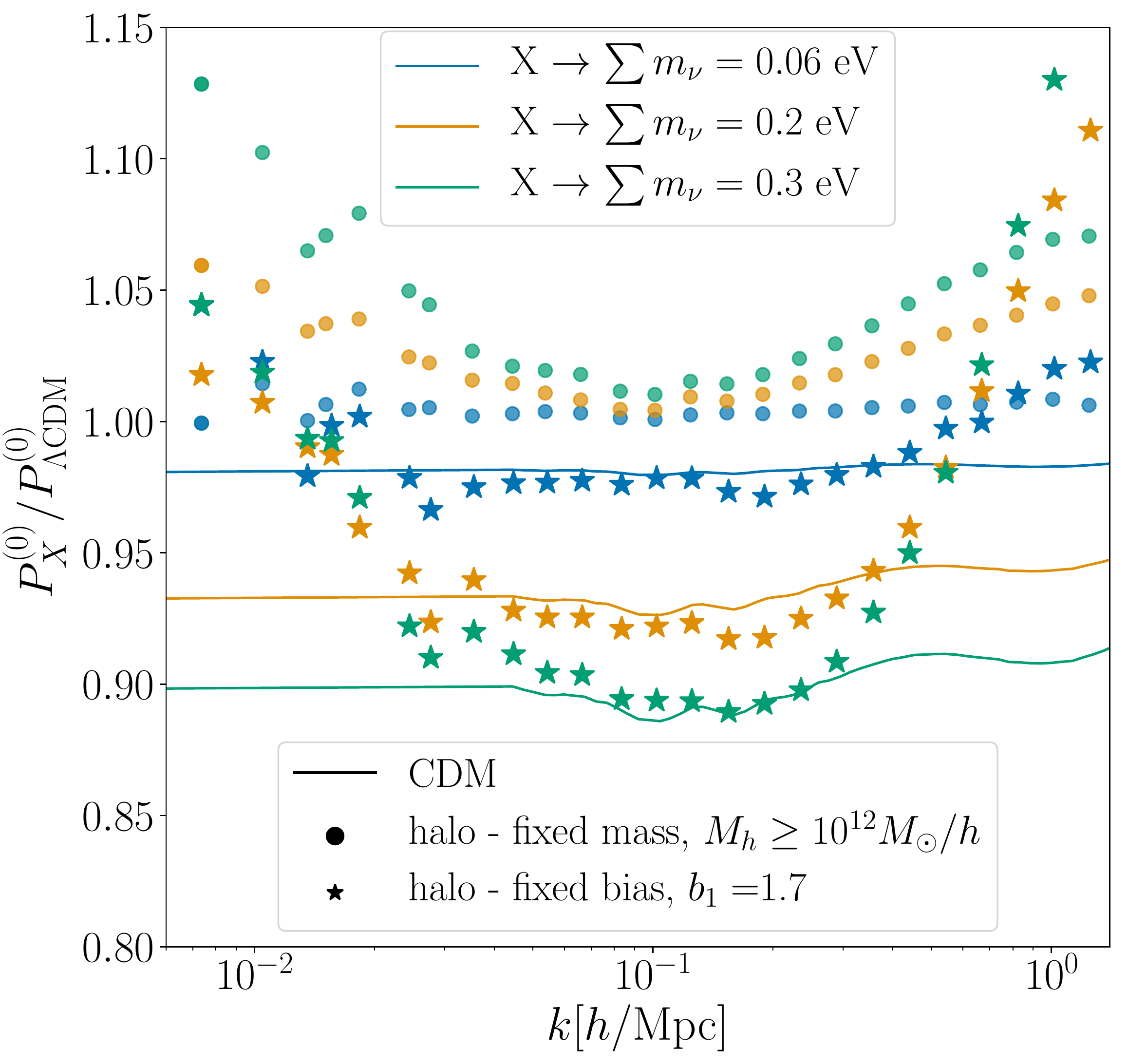} }}%
        \caption{Same as Fig.~\ref{kessence_pow}, but comparing models with different neutrino masses to $\Lambda$CDM (where neutrinos are massless). Only a halo selection at fixed bias (stars) recovers the monopole power ratio of matter (solid lines) on intermediate scales. Bias values and mass thresholds are chosen in a different way to in the $k$-essence comparison because of the better mass resolution.}
    \label{nu_pow}%
\end{figure}

\begin{figure}[tb]%
    \centering
    \hspace*{-1cm}  
   {{\includegraphics[scale=0.33]{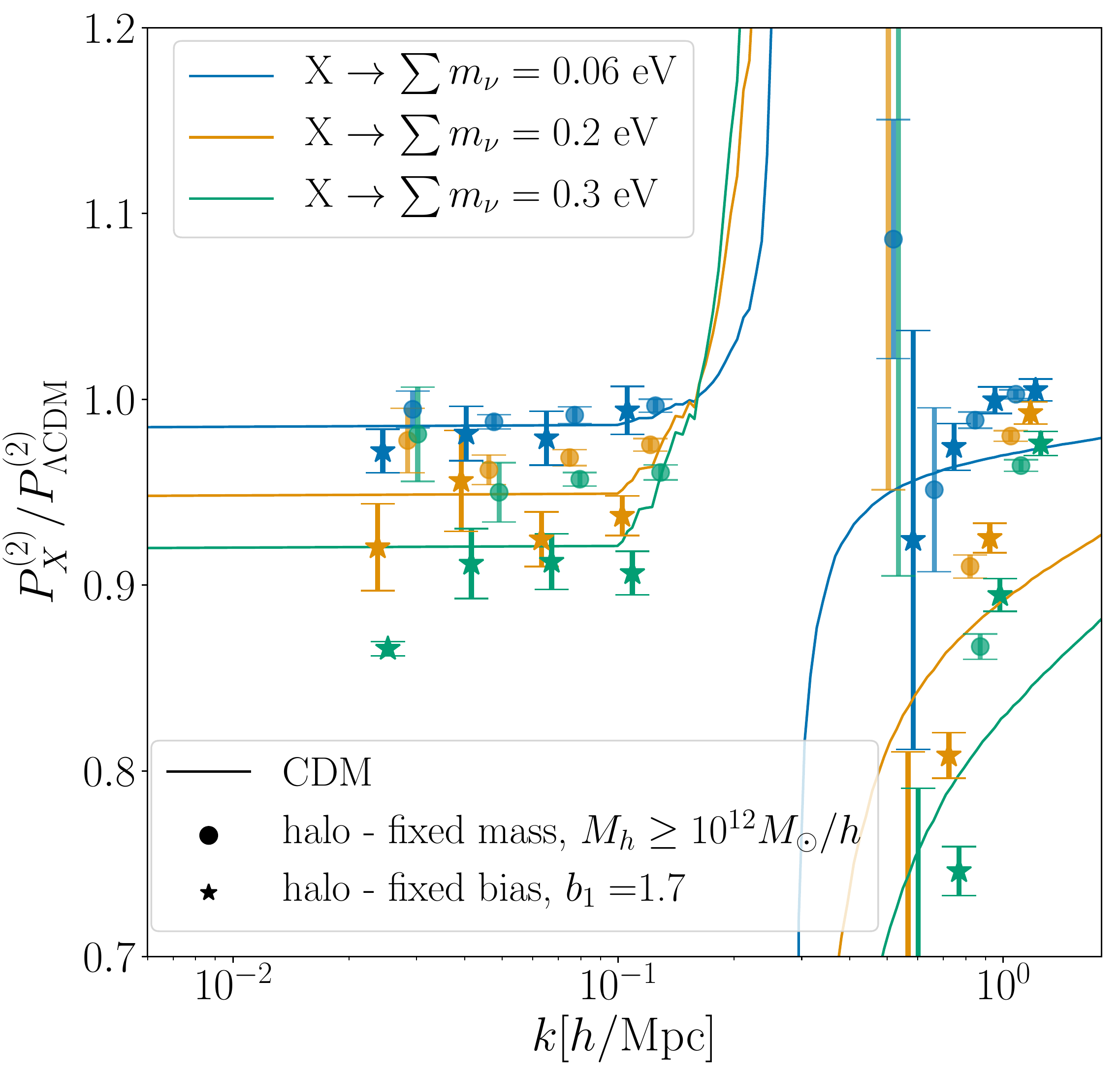} }}%
        \caption{  Ratio with respect to the massless case ($\Lambda$CDM) of the quadrupole of the halo power spectra (symbols with error bars) compared to the corresponding ratios for matter (solid lines) for different neutrino masses. %We show the massive neutrino with different total masses in different colors.
        The stars represent the result for the fixed bias selection, while the circles correspond to the selection at fixed mass threshold. The divergence at $k\simeq 0.3h^{-1}$Mpc is caused by a zero-crossing of the quadrupole.} 
    \label{nu_quadr}%
\end{figure}

\begin{figure}[tb]%
    \centering
    \hspace*{-1cm}  
   {{\includegraphics[scale=0.33]{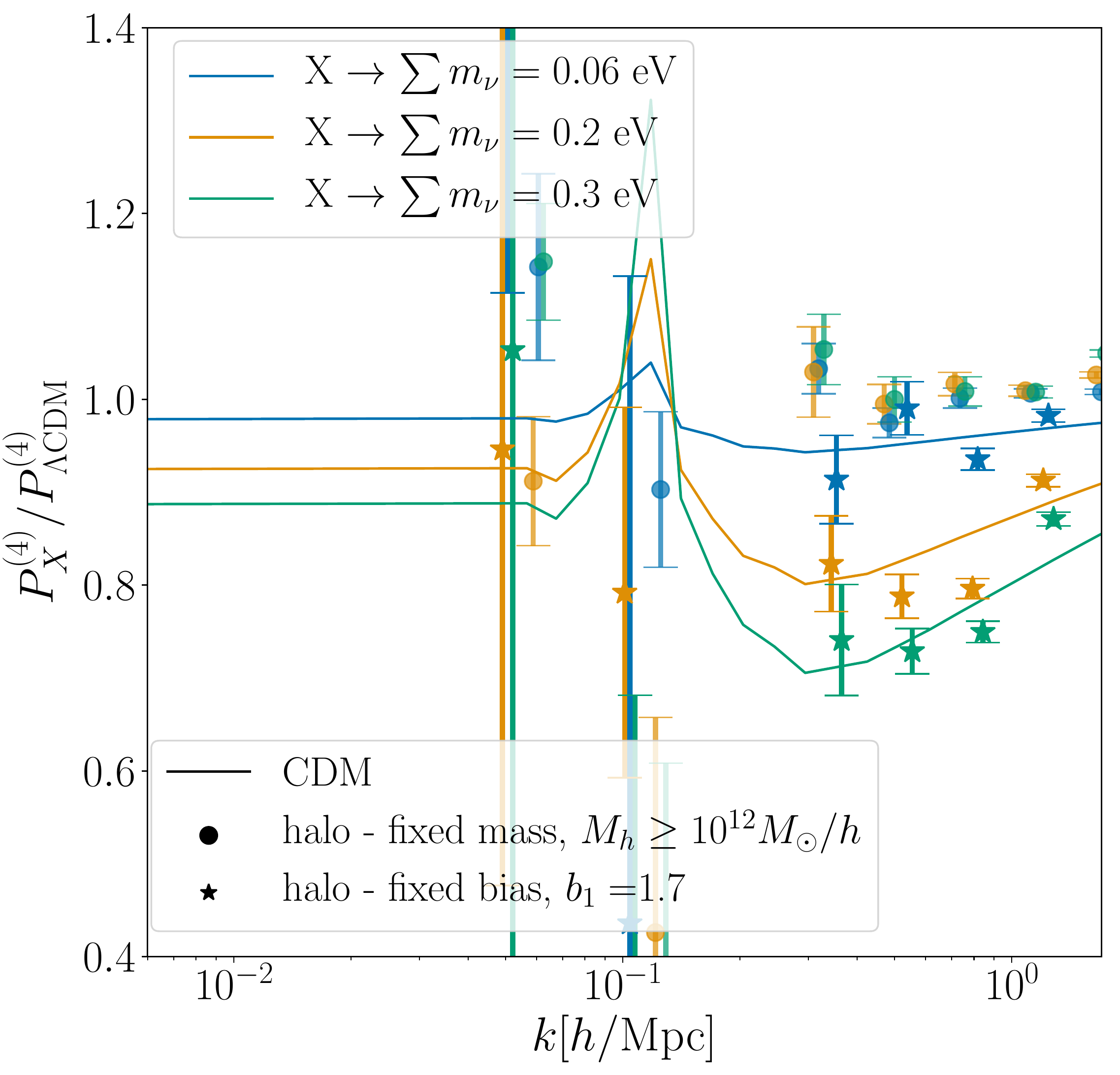} }}%
        \caption{ Same as Fig.~\ref{nu_quadr}, but for the hexadecapole of the power spectra.
        } 
    \label{nu_hexa}%
\end{figure}

In Fig.~\ref{nu_pow} we show the results for the monopole spectra of the massive neutrino models with the sum of neutrino masses $\sum m_{\nu} = 0.06$\,eV, $0.2$\,eV, and $0.3$\,eV in blue, orange, and green, respectively. Similar to the $k$-essence model, comparing halos selected by the same mass threshold within different models obscures the difference to $\Lambda$CDM at the intermediate scales, $0.03\,h/\mathrm{Mpc} \lesssim k \lesssim 0.3\,h/\mathrm{Mpc}$, where clustering measurements are most sensitive. While some differences are noticeable at larger and smaller scales, they bear no resemblance to the signatures seen in the matter, apparently even reversing some of the usual trends. These differences are somewhat difficult to interpret due to the relatively large contribution from RSD.
On the other hand, a comparison within the fixed bias halo samples recovers the signature of the matter power spectrum at 
the intermediate scales which is also where the bias is measured.
We also find the same behaviour in the higher moments of the power spectra as shown in Figs.~\ref{nu_quadr} and \ref{nu_hexa}. We see that comparing quadrupole and hexadecapole for the fixed bias halo selection recovers the 
signatures found in matter, although less clearly than in the monopole. This is mainly due to the larger uncertainties in the measurements of these higher moments which have significantly smaller amplitudes. On the other hand, we do not see any significant effect from different neutrino masses when we consider a fixed mass selection in different theories. 
It is also interesting to note that the ratios of the quadrupoles and hexadecapoles at fixed bias value agree relatively well with the ones of matter 
even in the nonlinear regime, $k>0.2h/$Mpc, where the corresponding monopole ratios deviate.

%%% monopole_nu

%%% Quadrupole

%%% Hexadecapole
 
From the bias definition in Eq.\ \eqref{bias_def} we can understand why comparing fixed bias halo samples in different models can recover the behaviour of matter at least for the scales on which the bias is linear. By selecting halo samples in $\Lambda$CDM and in model $X$ at the same bias value, we can write
\be \label{eq:3}
\frac{P^{\Lambda {\rm{CDM}}}_{hh}(k)}{P^{\Lambda {\rm{CDM}}}_{mm}(k)} = \frac{P^{X}_{hh}(k)}{P^{X}_{mm}(k)} \quad \Rightarrow \quad \frac{P^{X}_{hh}(k)}{P^{\Lambda {\rm{CDM}}}_{hh}(k)} =\frac{P^{X}_{mm}(k)}{P^{\Lambda {\rm{CDM}}}_{mm}(k)}.
\ee
This relation is valid on scales where the linear bias model is a good description in both cosmologies, and it %is exactly
can explain the results seen in Figs.~\ref{kessence_pow} and \ref{nu_pow} at linear and mildly non-linear scales, except that they show the monopole in redshift space which includes RSD.
In linear theory, the RSD can be modelled following \citet{Kaiser:1987qv}, leading to the well-known expressions for the multipoles of the power spectrum
\begin{eqnarray}
 P^{(0)}(k) &=& \left(1 + \frac{2}{3} \beta + \frac{1}{5} \beta^2\right) P(k)\,,\\
 P^{(2)}(k) &=& \left(\frac{4}{3} \beta + \frac{4}{7} \beta^2\right) P(k)\,,\\
 P^{(4)}(k) &=& \frac{8}{35} \beta^2 P(k)\,,
\end{eqnarray}
where $\beta = f / b_1$ and $f$ is the logarithmic growth rate in linear theory. Having fixed $b_1$ we can see from Eq.~\eqref{eq:3} that the halo power spectrum ratio would deviate from the matter power spectrum ratio by less than $\frac{2}{3} \Delta f$ for the monopole, by less than $\frac{3}{7} \Delta f$ for the quadrupole, and the hexadecapole ratio remains unaffected. Here, $\Delta f$ is the change in the growth rate between the two cosmologies. Large changes in the growth rate might therefore affect the monopole and quadrupole ratios, but for the models studied here we have $\Delta f$ of approximately a few percent at most, and therefore smaller than the typical change of bias when considering a fixed type of tracer. It should also be noted that the linear RSD model is not very accurate, and becomes completely useless at short distance scales where the so-called fingers-of-God effect becomes relevant. It is therefore interesting to see our results play out on those non-linear scales.

   \section{Conclusions}
   % The importance, future surveys
   Near-future LSS surveys will put very tight bounds on cosmological parameters. They will measure the sum of the neutrino masses and will help us to understand the nature of dark energy.
   To exploit the full potential of these surveys in better constraining our theories using  biased tracers, we need to understand how a signal in the matter distribution is translated into the halo and the galaxy clustering statistics.

   % In this work:
   In this work, we show for the first time that a naive selection of the halo sample based on a fixed mass threshold within different theories leads to the almost complete removal of the signature in the halo power spectrum. For this purpose, we used suites of large-scale cosmological simulations for two different models, namely $k$-essence and massive neutrinos. We demonstrate that, even in the higher moments of the power spectra, one sees a removal of signal for a fixed mass selection. On the other hand, when considering a selection at fixed value of the linear bias, the signal is recovered in the halo power spectra. The simple argument leading to this conclusion is summarised in Eq.~\eqref{eq:3} and we expect it to also hold for other model comparisons.
   On non-linear scales, the relation between the matter power spectrum and the halo power spectrum is difficult to model and is beyond the scope of this paper.
   
The results of this paper also show the importance of modelling or measuring the bias accurately.
The bias can in principle be measured by correlating weak lensing shear (which depends on total matter) and clustering statistics of the tracers, but the current sensitivity is still rather poor.
Using a measured bias, we can construct the tracer catalogue accordingly such that we see the expected effect in the tracer clustering.
 
\begin{acknowledgements}
We  thank Vincent Desjacques for helpful discussions. This work is supported by a grant from the Swiss National Supercomputing Centre
(CSCS) under project ID s1051. We have benefited from the Baobab and Yggdrasil clusters at the University of Geneva. Part of the numerical calculations were performed on the computational facilities at University of Oslo. FH is supported by the Overseas Research Fellowship from the Research Council of Norway. JA, RD and MK acknowledge financial support from the Swiss National Science Foundation.       
\end{acknowledgements}

%-------------------------------------------------------------------
% \clearpage
%\bibliographystyle{abbrv}
%\bibliography{bibliography.bib}
\bibliographystyle{aa}
\bibliography{bibliography}

\end{document}